\documentclass[floatfix,prl,twocolumn,showpacs,amsmath,amssymb,nofootinbib,letterpaper,groupaddresses,superscriptaddress]{revtex4}

\usepackage{amsmath}
\usepackage{amsthm}
\usepackage{amsfonts}
\usepackage{amssymb}
\usepackage{graphicx,color}
\usepackage{bbm}
\usepackage{textcomp}

\usepackage[usenames,dvipsnames]{xcolor}

\newcommand{\Tr}{\mathrm{Tr}}

\newcommand{\bra}[1]{\langle #1 |}
\newcommand{\ket}[1]{| #1 \rangle}
\newcommand{\bk}[2]{\langle #1 | #2 \rangle}

\newcommand{\supp}{\mathrm{supp}}

\newcommand{\kb}[2]{| #1 \rangle \langle #2 |}

\newtheorem{lemma}{Lemma}
\newtheorem{lemma1}{Lemma}

\begin{document}

\title{Security of two-way quantum key distribution}

\author{Normand J. Beaudry}
\affiliation{Institute for Theoretical Physics, ETH Zurich, 8093 Zurich, Switzerland}
\author{Marco Lucamarini}
\affiliation{Cambridge Research Laboratory, Toshiba Research Europe Ltd., 208 Cambridge Science Park, Milton Road, Cambridge, CB4 0GZ, United Kingdom}
\author{Stefano Mancini}
\affiliation{School of Science and Technology, University of Camerino, Camerino 62032, Italy}
\author{Renato Renner}
\affiliation{Institute for Theoretical Physics, ETH Zurich, 8093 Zurich, Switzerland}

\begin{abstract}
Quantum key distribution protocols typically make use of a one-way quantum channel to distribute a shared secret string to two distant users. However, protocols exploiting a two-way quantum channel have been proposed as an alternative route to the same goal, with the potential advantage of outperforming one-way protocols. Here we provide a strategy to prove security for two-way quantum key distribution protocols against the most general quantum attack possible by an eavesdropper. We utilize an entropic uncertainty relation, and only a few assumptions need to be made about the devices used in the protocol. We also show that a two-way protocol can outperform comparable one-way protocols.
\end{abstract}


\maketitle

\section{Introduction}

Quantum key distribution (QKD) research has been primarily focused on one-way protocols: one party, Alice, prepares states, sends them through an insecure quantum channel, and then another party, Bob, does a measurement \cite{scarani09,gisin02}. However, in the last decade, two-way protocols have been proposed where Bob prepares states, sends them to Alice through an insecure quantum channel, Alice does an encoding on the states, sends them backwards through the same quantum channel, and then Bob performs a measurement \cite{cai04a,cai04b,deng04,bostrom02,lucamarini05}. Paradigmatic examples of these kind of protocols are the so-called ``Ping-Pong" protocol \cite{bostrom02} and the LM05 protocol \cite{lucamarini05}. The former uses entangled states, while the latter uses non-orthogonal states. They have also been experimentally realized \cite{cere06,kumar08,ostermeyer08,khir12}.

It is not yet clear what the full potential of two-way protocols is, but there are at least several reasons why they are interesting. One motivation is that some two-way protocols are deterministic, that is, they do not require any sifting of the raw keys generated due to a mismatch of basis choices. For example, the LM05 protocol \cite{lucamarini05} has this advantage. The Ping-Pong protocol, which is based on super dense coding (SDC) \cite{bennett92b}, has no basis choices and therefore is also deterministic. Moreover, this protocol is conceptually interesting, as SDC can be turned into a QKD protocol.

One implementation of two-way protocols is to use polarization encoding of photons in fiber optics. The polarization drift caused by the fiber then needs to be actively corrected \cite{yuan05,dixon10,marand95}. However, if signals are sent backwards through the same channel, then the polarization drift is passively corrected by the use of a Faraday mirror at Alice's side. This means that there may be experimental situations in which one-way QKD does not succeed because the error rate is too high but two-way QKD may still be possible. One QKD protocol that exploits this fact is the ``Plug \& Play" BB84 protocol \cite{bb84,ribordy98}. This implementation ideally yields one raw key bit for each qubit signal sent from Bob to Alice and then from Alice to Bob. From SDC we know that two bits can be communicated by only sending one qubit in this manner. Therefore, another motivation is that the key rate can be increased by using the SDC protocol instead of Plug \& Play BB84 by using the same channel resources and without the need of higher dimensional states or more complicated measurements.

A major difficulty when studying the security of two-way protocols is that the eavesdropper, Eve, can attack each signal twice: once on the way from Bob to Alice, and later on its way back from Alice to Bob. This gives her more strategies than in a one-way QKD protocol. In fact, the Ping-Pong protocol has been shown to be insecure \cite{wojcik03,cai03}, while recently the LM05 protocol was proven secure, but by assuming the use of qubits and the full characterization of all of the devices \cite{lu11,fung12a}. Also, the Plug \& Play protocol was proven secure \cite{zhao08,zhao10} but by using strong assumptions (e.g.~an intensity monitor, phase randomizer, and attenuator are required, and all devices, except the source, are fully characterized).

Unlike these previous approaches, we propose a general security proof strategy through which devices used in the protocol only need to be characterized by a few assumptions. Our assumptions are on the same level as one-way QKD security proofs that use uncertainty and complimentarity, such as the proofs by Mayers \cite{mayers96} and Koashi \cite{koashi09}. This is in contrast to device-independent security proofs where no assumptions are made about devices used in the protocol. However, they require loophole free Bell tests, which are not possible with current technology \cite{barrett05,barrett12,vazirani12}. Our assumptions lie between the device-dependent scenario, where devices are completely characterized, and the device-independent scenario. For example, a device may be characterised by a single constant that can be experimentally bounded.

Our proof strategy consists of two main steps. First we show how to ``purify'' prepare and measure protocols into entanglement based protocols. Second, we apply the entropic uncertainty relation proposed as a tool for security proofs of one-way protocols to the purified protocols \cite{berta10}. An entanglement-based or purified protocol is one where Eve prepares a state, sends a part of the state to Alice and another part to Bob, and then Alice and Bob perform measurements. The uncertainty relation we use states \cite{berta10}: given a tri-partite quantum state $\rho_{ABE}$ and two measurements on system $A$, $F_{X}$ and $F_{Z}$, described by elements of a positive-operator valued measure (POVM) $\{F^{i}_{X}\}_{i}$ and $\{F^{i}_{Z}\}_{i}$ with classical outcomes $X$ and $Z$ respectively, then
\begin{equation} \label{eq:ur}
H(Z|B)+H(X|E) \geq \log_2 1/\gamma,
\end{equation}
where $H(A|B)$ is the conditional von Neumann entropy, and $\gamma:=\max_{i,j}\|\sqrt{F^{i}_{X}}\sqrt{F^{j}_{Z}}\|^{2}_{\infty}$ (which we call the overlap between the measurements $F_X$ and $F_Z$). Given an operator $F$ acting on a Hilbert space $\mathcal{H}$ such that $F\geq 0$, then $\|F\|_{\infty}:=\max\{\bra{\phi}F\ket{\phi} : \phi\in \mathcal{H},\bk{\phi}{\phi}=1\}$ is the operator norm on positive operators. Using this uncertainty relation and the Devetak-Winter security bound \cite{devetak05}, we demonstrate how to prove security against the most general type of attacks for two-way protocols.

Actually we use this method to prove security for two example protocols: a super dense coding (SDC) protocol similar to the Ping-Pong protocol \cite{bostrom02} and a protocol similar to LM05 (which we will also refer to as LM05) \cite{lucamarini05}. For the LM05 protocol we show an improvement of the key rate of \cite{lu11}. Furthermore, we provide a comparison among relevant two-way and one-way protocols showing that the former can outperform the latter.

Our proof  clarifies the analysis of two-way QKD protocols and provides an important step towards device-independent security of quantum cryptography in this framework. In addition, our results illustrate that the uncertainty relation Eq.~\ref{eq:ur} can be useful to prove security of QKD protocols other than BB84.

We proceed by first defining the SDC and LM05 protocols in the scenario where only qubits are used. Second, we describe purified versions of these protocols in order apply the uncertainty relation Eq.~\ref{eq:ur}. Third, we list the assumptions that are needed for the application of this security proof to implementations of these protocols. Fourth, we prove the security of the protocols. Lastly, we compare the key rates to different implementations of the BB84 protocol.

\section{Protocol Definitions}

In the descriptions of the SDC and LM05 protocols below we assume that the states are deterministically prepared and all devices are completely characterised. This is for the ease of describing the protocols and this assumption is not be necessary for the security proofs.

There are some similarities between both protocols: they have two quantum channels between Alice and Bob, $Q_1$ and $Q_2$, which can be attacked by the eavesdropper, Eve, using any strategy allowed by quantum mechanics. Also, Alice and Bob will be performing some $X$- and $Z$-basis measurements. These refer to the projections onto the eigenvectors of the Pauli operators $\sigma_X$ and $\sigma_Z$ respectively. In addition, Alice and Bob will do parameter estimation, error correction, and privacy amplification on their raw data after the steps outlined below. They abort their protocol if during parameter estimation they find that one of their relevant error rates is beyond a certain threshold.

\subsection{Qubit SDC Protocol}

\begin{figure}[h!]
  \includegraphics[width=8cm]{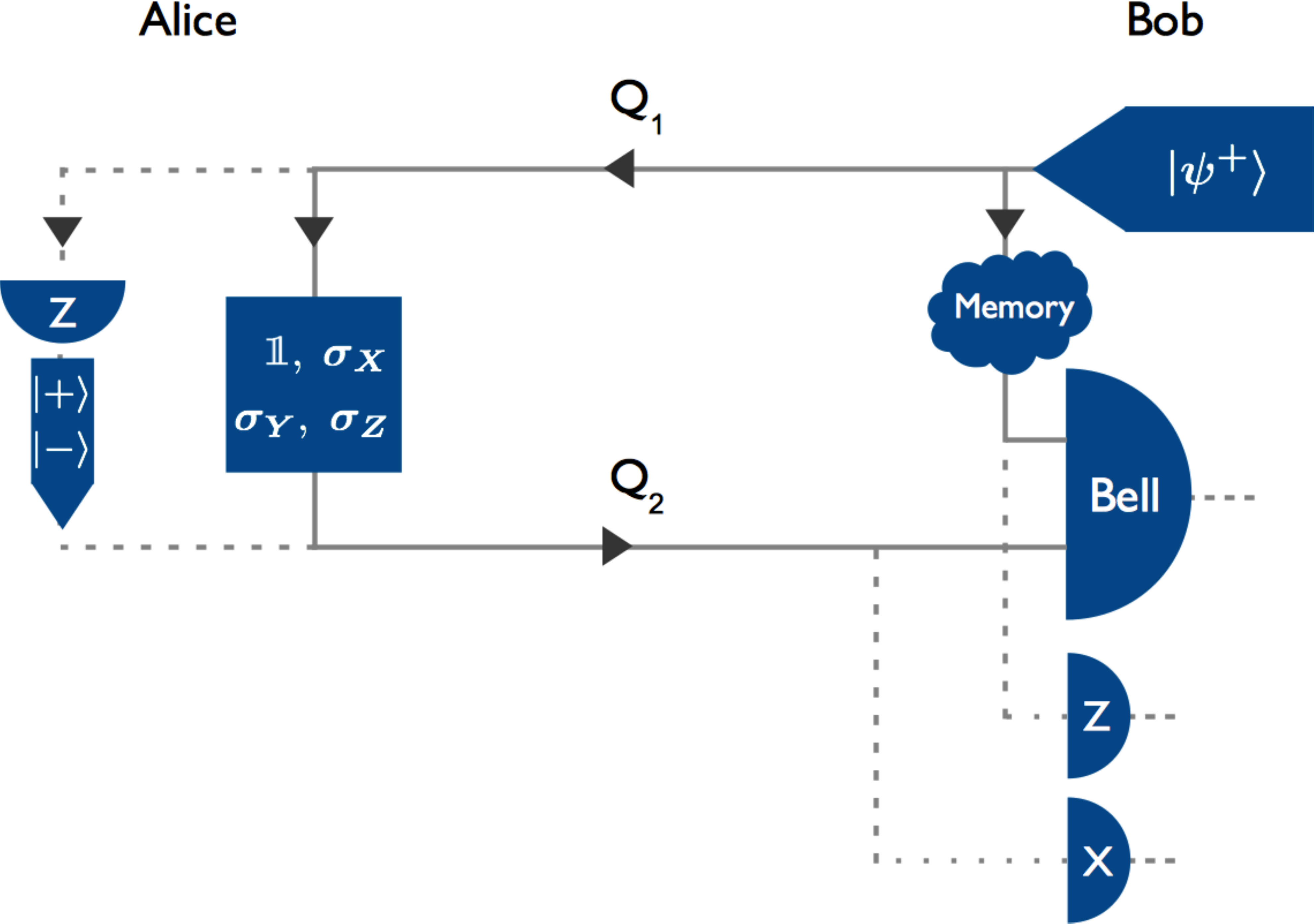}\\
  \caption{Depiction of the ideal qubit SDC protocol. Bob prepares the Bell state $\ket{\psi^+}$, and keeps half of it in a quantum memory. He sends the other half to Alice who either does an encoding (with probability $c$, solid line) or does a $Z$-basis measurement which indicates whether she prepares $\ket{+}$ or $\ket{-}$ (with probability $1-c$, dashed line). Alice sends this state to Bob who does a Bell measurement with his stored qubit and the qubit from Alice (with probability $c$, solid line) or does a $Z\otimes X$-basis measurement (with probability $1-c$, dashed line).}\label{fig:SDC}
  \vspace{-0.5cm}
\end{figure}

{\bf Bob's preparation}: Bob prepares a maximally entangled state $\ket{\psi^{+}}=1/\sqrt{2}(\ket{00}+\ket{11})$ and keeps one half of it in a quantum memory. He sends the other half to Alice through channel $Q_1$ (see Fig.~\ref{fig:SDC}).

{\bf Alice}: With probability $c\approx 1$ Alice applies one of the four Pauli operators $\mathbbm{1},\sigma_{X},\sigma_{Y},\sigma_{Z}$ (choosing each with probability $1/4$) to the state from the channel $Q_1$. She records her choice by storing two classical bits: $00,10,11,01$, respectivly. Alice then sends this state into channel $Q_2$ back to Bob. With probability $1-c$ Alice measures the state from channel $Q_1$ in the $Z$-basis. She then prepares $\ket{+}$ with probability $1/2$ or $\ket{-}$ with probability $1/2$, where $\ket{\pm}:=1/\sqrt{2}(\ket{0}\pm\ket{1})$, and sends it into channel $Q_2$ to Bob.

{\bf Bob's measurement}: With probability $c$ Bob performs a Bell measurement jointly on his stored qubit and his received qubit from the channel $Q_2$. He gets possible outcomes $\ket{\psi^{+}},\ket{\psi^{-}},\ket{\phi^{+}},\ket{\phi^{-}}$, and then will store the bits $00,01,10,11$ respectively. With probability $1-c$ he measures his stored qubit in the $Z$-basis and his received qubit in the $X$-basis.

{\bf Post-processing}: Alice and Bob repeat the above procedure $N$ times. Their raw key is the concatenation of all of their two-bit strings together respectively. Alice publicly announces which signals she encoded and which signals she measured.

\subsection{Qubit LM05 Protocol}

\begin{figure}[ht]
\begin{center}
  \includegraphics[width=8cm]{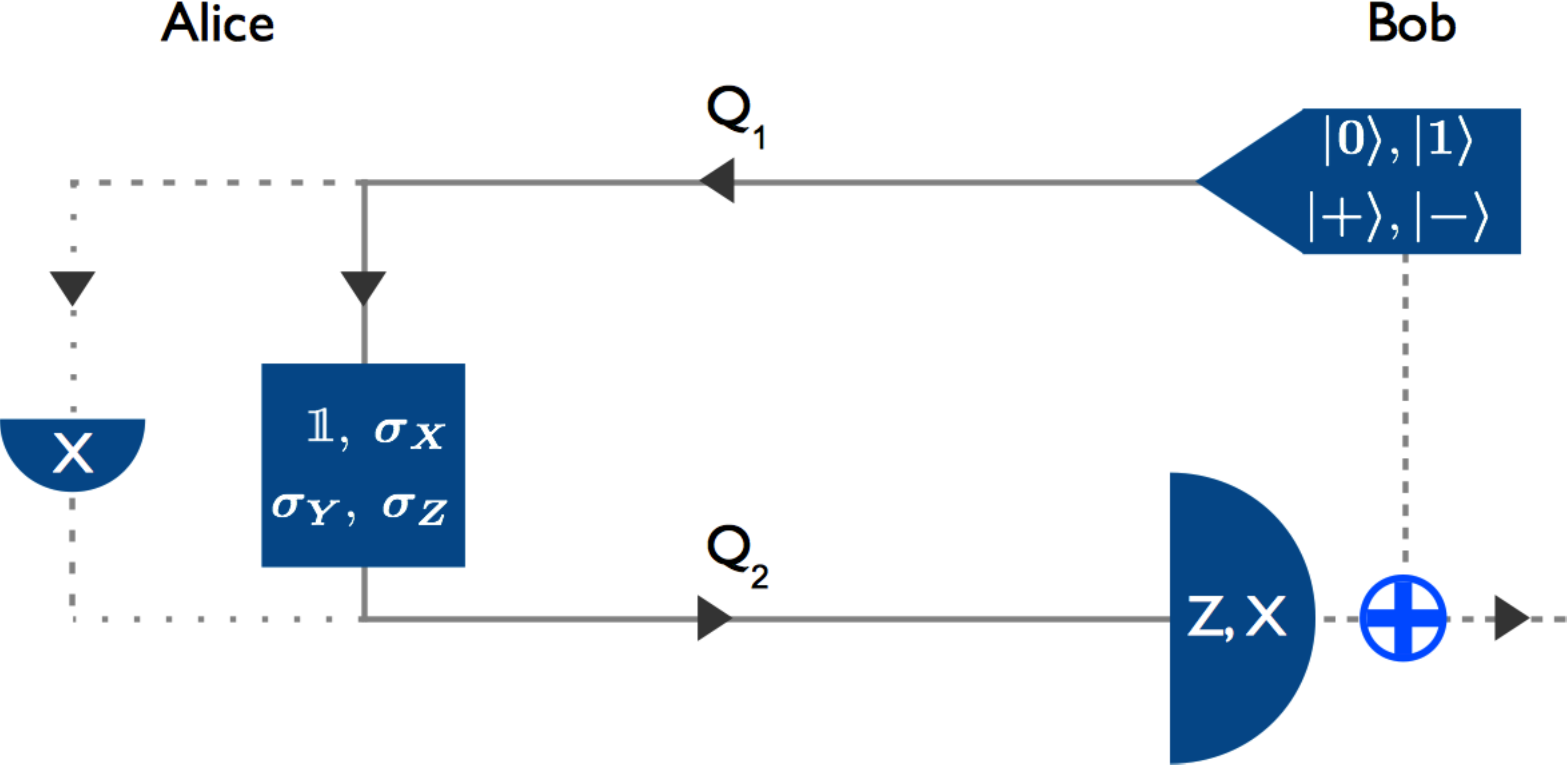}\\
  \caption{Version 1 of the ideal qubit LM05 protocol. Bob prepares one of the four BB84 states and sends it Alice. Alice performs an encoding (with probability $c$, solid line), or a measurement in the $X$-basis followed by the sending of the post-measurement state (with probability $1-c$, dashed line). Bob performs a measurement in the $Z$- or $X$-basis whenever he prepared states in the $Z$- or $X$-basis respectively. Bob then does an XOR of his measured bit and his preparation bit corresponding to his prepared state.}\label{fig:lm05}
\end{center}
\end{figure}

In the LM05 protocol \cite{lucamarini05} Alice and Bob will have the choice to perform either an $X$- or $Z$-basis measurement. We use a parameter $p$ to denote the probability that Alice and Bob choose the $Z$-basis, so $1-p$ is the probability that they choose the $X$-basis. We will consider two possible versions of the protocol for the simplicity of presentation. Version $1$ is when $p\approx 1$, and Alice and Bob will only use their $X$-basis measurement for parameter estimation (see Fig.~\ref{fig:lm05}). Version $2$ is when $p=1/2$ and then they will use both $X$- and $Z$-basis measurements for parameter estimation and key generation. Note that the choice of $p$ will not affect the key rate.

{\bf Bob's preparation}: Bob prepares one of the four states $\ket{0},\ket{1},\ket{+},\ket{-}$. He chooses $\ket{0}$ or $\ket{1}$ each with probability $p/2$, and $\ket{+}$ or $\ket{-}$ each with probability $(1-p)/2$. When he picks either $\ket{0}$ or $\ket{+}$ he classically stores a $0$, when he picks either $\ket{1}$ or $\ket{-}$, he stores a $1$ (we refer to this bit as the preparation bit). Bob also stores the basis the state is in. He sends the state to Alice through channel $Q_1$ (see Fig.~\ref{fig:lm05}).

{\bf Alice}: With probability $c\approx 1$ Alice applies one of $\mathbbm{1},\sigma_{X},\sigma_{Y},\sigma_{Z}$ (choosing each with probability $1/4$) to the state from the channel $Q_1$. She records her choice of encoding. With probability $1-c$ she applies a $X$-basis measurement (Version $1$), or randomly chooses either an $X$- or $Z$-basis measurement (Version 2). Alice then takes the post-measurement state (when a measurement was performed) or the encoded state (where a Pauli-operator was applied) and sends it to Bob through channel $Q_2$.

{\bf Bob's measurement}: Bob does a measurement in the same basis he prepared his state in: if he prepared $\ket{0}$ or $\ket{1}$ then he measures in the $Z$-basis, if he prepared $\ket{+}$ or $\ket{-}$ then he measures in the $X$-basis.

{\bf Post-processing}: Alice and Bob repeat this procedure $N$ times. If they perform reverse reconciliation then Bob publicly reveals which basis he used for each signal, and Alice reveals which signals she measured and which she encoded. In Version 2 Alice also reveals which basis she measured in for each signal and then Alice and Bob discard their measurement results wherever they measure in different bases. Bob's raw key is the result of the XOR of his measurement outcomes with his preparation bits. Alice's raw key is made up of one of the two classical bits $00,10,11,01$ corresponding to the encodings $\mathbbm{1},\sigma_{X},\sigma_{Y},\sigma_{Z}$, respectively. Whenever Bob measured in the $Z$-basis Alice keeps her first bit, and when Bob measured in the $X$-basis Alice keeps the second bit.

In direct reconciliation, Bob does not reveal his basis choice and instead Alice reveals whether she applied one of the encodings from the set $S_{0}:=\{\mathbbm{1},\sigma_{Y}\}$ or the set $S_{1}:=\{\sigma_{X},\sigma_{Z}\}$. Alice corresponds the encodings $\mathbbm{1},\sigma_{X},\sigma_{Y},\sigma_{Z}$ with the bits $0,1,1,0$ respectively. Bob then needs to flip his raw bit for each signal that he used the $X$-basis and Alice announces she applied an encoding from $S_1$. 

We now purify these two prepare and measure QKD protocols by showing that they are equivalent to protocols that start with entangled states distributed by Eve followed by measurements by Alice and Bob.

\section{Purified Protocols}

We introduce two purified protocols that are structured such that a pure state is shared between Alice, Bob, and Eve; and then Alice and Bob perform measurements on this state. These purified protocols are equivalent to the prepare and measure protocols described above. However, less assumptions are needed about the devices used. In the next section we will explain exactly which assumptions about the devices in the prepare and measure protocol are necessary in order to apply our security proofs. Afterwards we will prove the security of the purified protocols.

\begin{figure}[h]
\begin{center}
  \includegraphics[width=8cm]{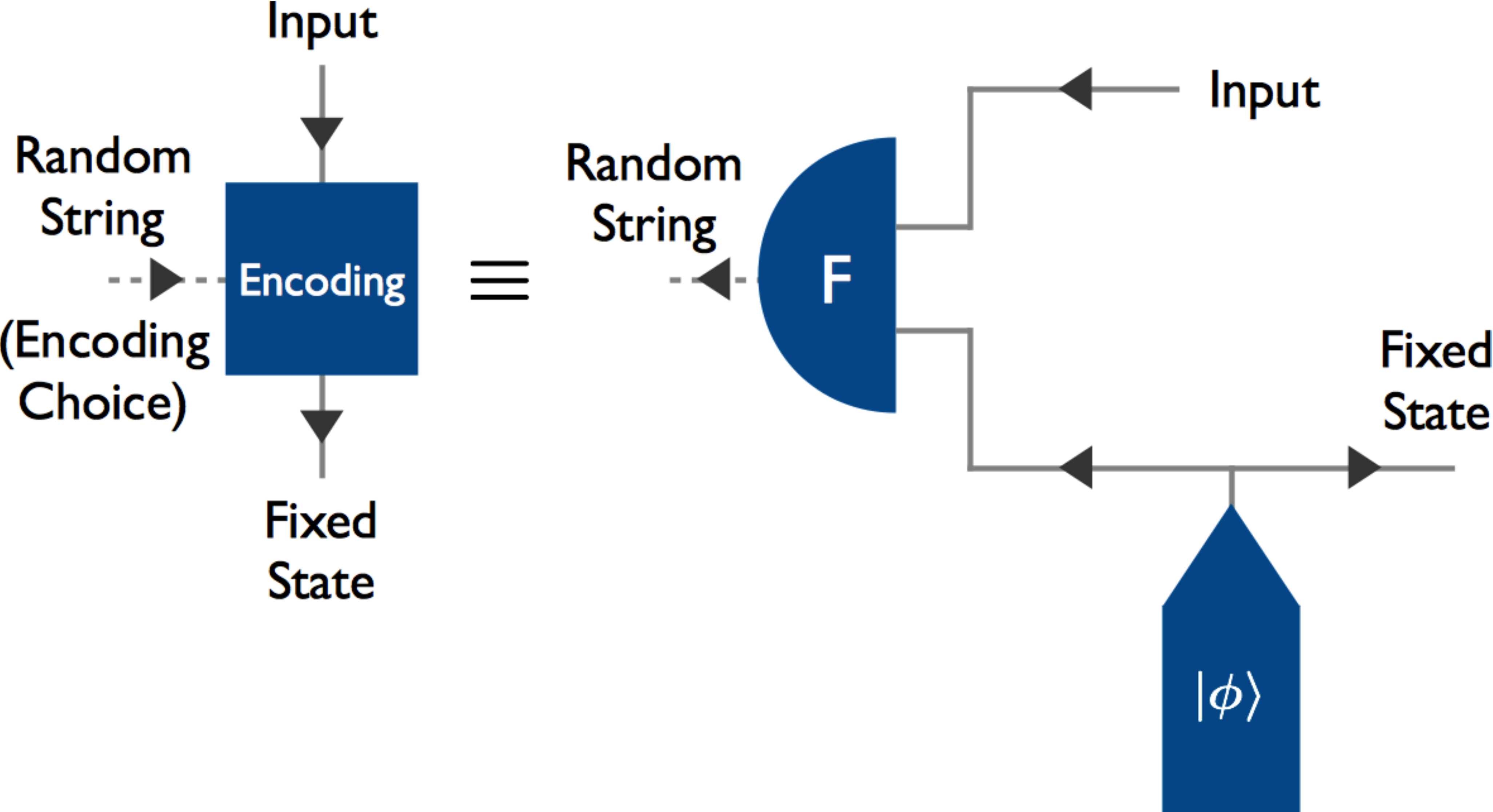}\\
  \caption{A depiction of Lemma 1. An encoding that takes a quantum state and a random string as input, which is used to choose which encoding to perform, always outputs a fixed state (averaged over all encoding choices). This encoding is equivalent to the scenario where a measurement, $F$, acts jointly on the same quantum state input as the encoding and half of a bipartite pure state $\ket{\phi}$. The output of the measurement is a random string, and the other half of $\ket{\phi}$ is then the same fixed state output from the encoding, averaged over all measurement outcomes of $F$.}
  \label{fig:Lemma1}
\end{center}
\vspace{-0.5cm}
\end{figure}

We can purify Alice's encoding operation in both protocols by finding an equivalence to a POVM acting on the input of the encoding and half of a pure state such that the other half of the pure state is the same as the output from Alice's encoding (see Fig.~\ref{fig:Lemma1}). In addition, the outcome of the POVM is two random bits independent of the input and therefore is the same as Alice's choice of encoding operation using a random string. We use the following lemma to achieve this equivalence (see the appendix for the proof).
For the lemma, we define the set of normalized positive semi-definite operators on a Hilbert space $\mathcal{H}: S(\mathcal{H}) := \{\tau\in \mathcal{P}(\mathcal{H}):\Tr(\tau)=1\})$, where $\mathcal{P}(\mathcal{H})$ is the set of positive semi-definite operators on $\mathcal{H}$.
\begin{lemma}\label{lemma:POVMCPTP1} Let $\{\mathcal{E}_{i}\}_{i=1..n}$ be a set of $n$ completely positive trace-preserving maps from Hilbert space $\mathcal{H}_A$ to Hilbert space $\mathcal{H}_D$ and $\sigma_{D}\in S(\mathcal{H}_{D})$ be a fixed density operator on $\mathcal{H}_{D}$ such that
\begin{equation} \label{eq:indepE}
1/n\sum_{i=1}^{n}\mathcal{E}_{i}(\rho_{A})=\sigma_{D}\;\forall \rho_{A}\in S(\mathcal{H}_{A}).
\end{equation}
Then there exists a fixed pure state $\ket{\phi}_{CD}$  in $\mathcal{H}_{CD}:=\mathcal{H}_{C}\otimes\mathcal{H}_{D}$, where $\dim \mathcal{H}_{C} = \dim\mathcal{H}_{D}$, and a complete set of POVM elements $\{F^{i}_{AC}\}_{i=1..n}$ on $\mathcal{H}_{AC}$ $($so $\sum_{i}F^{i}_{AC}=\mathbbm{1}_{AC})$, such that $\forall i,\forall\rho_{A}\in S(\mathcal{H}_{A})$ we have
\begin{equation}\label{eq:equiv1}
n\Tr_{AC}\left(F_{AC}^{i}\rho_{A}\otimes\ket{\phi}_{CD}\bra{\phi}\right)=\mathcal{E}_{i}(\rho_{A}).
\end{equation}
\end{lemma}
In the prepare and measure protocol there were four encodings that Alice could do, and therefore $n=4$ for the application of this lemma. Now that the encoding is purified, we purify all of the preparations of both protocols.

In the perfect qubit version of both protocols, Alice and Bob's preparations are equivalent to starting with a maximally entangled state $\ket{\psi^+}=1/\sqrt{2}(\ket{00}+\ket{11})$ followed by a $Z$- or $X$-basis measurement (or probabilistic distribution over the choice of the two measurements). More generally, if we only assume that the preparations are of qubits then they are equivalent to the maximally entangled state $\ket{\psi^+}$, followed by a measurement on one of the two qubits. The non-measured qubit is then the same as the prepared qubit \cite{tomamichel12a}.
It only makes Eve more powerful to prepare both entangled states from Alice's encoding and from the purifications of the preparations, so we can let her prepare these states.

Now the protocols can be described as follows. Eve prepares a state $\rho_{ABE}$ and sends $A$ to Alice and $B$ to Bob, and keeps part $E$. Alice and Bob perform one of two measurements on each of their systems. Then Alice and Bob do post-processing as in the prepare and measure protocol. Before we prove the purified protocol's security, we outline which assumptions we will make for the security proof to hold.

\section{Assumptions}
 
 To specify exactly when our security proofs will apply, we explicitly state the assumptions we will make about the devices used in the protocols. Afterward we will discuss how these assumptions can be justified.
 
\begin{enumerate}
\item The states prepared are qubits. This assumption can be avoided if instead an arbitrary bipartite state is prepared and a measurement is performed on one half of it, while sending the other half into the appropriate channel. It can be assumed that Eve holds a purification of this prepared bipartite state.
\item The output state of Alice's encoding operation is independent of the input state, when averaged over all encodings. Mathematically, this means that the encoding satisfies Eq.~\ref{eq:indepE} in Lemma~\ref{lemma:POVMCPTP1}.
\item Measurements detect each signal independently. This means that the POVM elements have an i.i.d. form.
\item Bob's devices (when reverse reconciliation is performed) or Alice's devices (when direct reconciliation is performed) are characterized by a constant overlap, $\gamma$, defined after Eq.~\ref{eq:ur}.
\item There are no losses in the channels or detectors.
\end{enumerate}
Assumption 1 is needed in order to purify the preparation process. Assumption 2 is needed in order for the encoding to have a purified form. Assumption 3 is needed so that we can analyze the measurement outcomes in an i.i.d.~way. Assumption 4 is needed to apply the uncertainty relation Eq.~\ref{eq:ur}. Assumption 5 is made for the convenience of the security analysis, but can be removed if Alice and Bob randomly assign a bit value for their measurement in a time where a signal was sent but they did not receive a measurement outcome.

Assumption 1 is valid if the preparations are done in a purified way. Assumptions 2 and 3 are idealized assumptions necessary for our security proof method. However, Assumption 3 is common for both device-independent and device-dependent security proofs \cite{barrett05,barrett12,vazirani12}. Assumption 2 requires knowledge about Alice's encoding device, and could be justified to a good approximation. Such assumptions are typical for device-dependent security proofs \cite{scarani09}. However, Assumption 2 is not necessary if the purified protocol is implemented directly.

The overlap $\gamma$ that characterises the devices in Assumption 4 cannot be obtained without a description of the POVMs for the measurements (if reverse reconciliation is performed), or a description of the encoding map (if direct reconciliation is performed). However, Alice and Bob can put a bound on a so-called `effective overlap': $\gamma^*$. This effective overlap is defined differently than the overlap in Eq.~1 (see Definition 7.2 in \cite{tomamichelthesis}), but it satisfies the same uncertainty relation:
\begin{equation}
H(Z|B)+H(X|E)\geq \log 1/\gamma^{*}.
\end{equation}
The effective overlap has the advantage that it can be upper bounded by measuring a CHSH value \cite{CHSH} if the measurements have binary outcomes. We now define the CHSH value and then we will describe the upper bound on the effective overlap. Given a bipartite system, let $M$ and $N$ be random variables representing the measurement outcomes from choosing one of two measurements randomly on one part of the system, and let $R$ and $S$ be random variables representing the measurement outcomes from choosing one of two measurements randomly on the other half of the system. Then the CHSH value, $\beta$, is defined as
\begin{align}
\beta:= 2(&\textrm{Pr}(M=R)+\textrm{Pr}(N=R)+\nonumber\\ &\textrm{Pr}(M=S)+\textrm{Pr}(N\neq S))-4.\nonumber
\end{align}
Since the LM05 protocol measurements and encoding choice are binary, the CHSH value provides an upper bound for the effective overlap. It is not yet known if the same relation between the effective overlap and a CHSH value holds for more than two measurement outcomes \cite{tomamichel13}. Therefore, it is not known how the effective overlap may be upper bounded by the CHSH value for the SDC protocol since there are four measurement outcomes and four encoding choices in this protocol.

For the LM05 protocol, as long as preparations are done in a purified way, Alice and Bob can run Version 2 of the LM05 QKD setup and find the CHSH value between Alice's encoding choice and Bob's measurement outcomes. More precisely, Bob's XOR of his $Z$- and $X$-basis measurement outcomes with his preparation bit define the random variables $R$ and $S$, while Alice's bit values that correspond to the encoding sets $S_0$ and $S_1$ define $M$ and $N$. Despite Alice not performing a measurement we can still define the CHSH value this way, since from Assumption 2 we are guaranteed that Alice's encoding corresponds to a measurement via Lemma~1, where her encoding choice corresponds to the measurement's outcomes.

The effective overlap is bounded by
\begin{equation}\label{eq:beta}
\gamma^* \leq \frac{1}{2} +\frac{\beta}{8}\sqrt{8-\beta^2}.
\end{equation}
Note that no additional devices are needed to put a bound on the effective overlap. Also, if desired, Alice or Bob (depending on whether direct or reverse reconciliation is performed) can measure the CHSH value by themselves by using another measurement device on their side. Alice can measure her CHSH value in the same way as Alice and Bob did it jointly. Bob can measure his CHSH value by running the purified QKD setup. In summary, Assumption 4 can be justified by an experimental test on the devices used in the LM05 protocol (see \cite{tomamichelthesis,tomamichel13} for more details).

For the security proofs below we fix the overlap in Assumption 4 for each protocol. The overlap for the SDC protocol, when reverse reconciliation is performed, between Bob's two measurements is assumed to be $1/4$ (which is true for the ideal Bell and $Z\otimes X$-basis measurements). For the LM05 protocol, where Bob's preparations are done with a bipartite state and a measurement, we assume that there are two measurements that have an overlap of $1/2$ with Bob's measurements followed by an XOR of the outcomes. The first is his measurement on half of his prepared pure state in the other basis and the second is his measurement in the other basis on channel $Q_2$. Note that this overlap occurs between the ideal $Z\otimes Z$-basis measurement followed by an XOR of the outcomes and the $X$-basis measurement on his prepared pure state and his $X$-basis measurement on the input from channel $Q_2$. In the case of direct reconciliation this assumption changes to the overlap between Alice's POVM associated with her encoding (via Lemma~1) and her measurement tensored with her purified preparation measurement. While we have made these rigid assumptions for the security proofs, we can relax the assumption that these overlap is exactly $1/2$ for the LM05 protocol and instead use the CHSH value bound on the effective overlap \cite{tomamichelthesis,tomamichel13}.

Assumption 5 is clearly not experimentally justifiable. However, it can be removed if, whenever there is a missing measurement outcome at Alice or Bob's detector, Alice and Bob randomly assign a bit value. The error rates will be increased, decreasing the key rate significantly. We leave a more detailed analysis of loss as future work, which could follow along similar lines as \cite{tomamichel12a}.

It is important to note that no assumptions are necessary about the Hilbert space that the signals of the protocols are in (except, possibly, qubit preparations). In addition, no assumptions need to be made about the internal structure of the measurements on each signal, descriptions of the preparations of bipartite states, or the quantum memory used in the SDC protocol.

\section{Security Proofs}

The security proofs of the purified protocols can be found via the Devetak-Winter rate \cite{devetak05}, followed  by the application of the uncertainty relation of Eq.~\ref{eq:ur} \cite{berta10}. The security proofs can then be applied to the non-purified SDC and LM05 protocols since they are equivalent to the purified protocols. This equivalence is guaranteed under Assumptions 1 and 2 of the previous section.

\subsection{SDC Protocol}

Now we define some states useful for the security proof.
The state that Alice, Bob, and Eve share after Alice and Bob have done their measurements is
\begin{equation}
\tau_{Z_{A}Z_{B}E} = F_{A}\otimes F_{B} (\rho_{ABE}),
\end{equation}
where $Z_{A}$ and $Z_{B}$ are the classical strings that result from Alice and Bob's measurements, $F_{A}$ and $F_{B}$, which are represented as completely positive trace preserving (CPTP) maps. We assume that the measurements $F_A$ and $F_B$ act independently on each signal, so that we can apply the uncertainty relation to each measurement independently. Using Assumption 3 from the previous section, the measurement's POVM elements have the form $\{\bigotimes_{j}F^{i_j}_{A}\}_{i}$ and $\{\bigotimes_{j}F^{i_j}_{B}\}_{i}$, where $i=i_{1}i_{2}i_{3}\dots$. We also define another state, $\xi$, where we only change Alice's measurement. This state has the important property that $H(Z_{B}|E)_{\tau}=H(Z_{B}|E)_{\xi}$. Intuitively this means that Eve's information about Bob's string does not depend on Alice's measurement. The state $\xi$ is defined as
\begin{equation}
\xi_{X_{A}Z_{B}E} = G_{A}\otimes F_{B} (\rho_{ABE}),
\end{equation}
where $X_A$ is the classical string output from the measurement $G_{A}$ on Alice's side. We do not characterize $G_{A}$. However, we assume that the POVM elements of $G_{A}$ are independent, and therefore have the form  $\{\bigotimes_{j}G^{i_j}_{A}\}_{i}$ (Assumption 3). We now define a third state that will be used for the application of the uncertainty relation \cite{berta10}:
\begin{equation}
\sigma_{X_{A}X_{B}E} = G_{A}\otimes G_{B} (\rho_{ABE}),
\end{equation}
where $G_{B}$ have POVM elements of the form $\{\bigotimes_{j}G^{i_j}_{B}\}_{i}$ (Assumption 3) and its classical output is denoted as $X_{B}$. In addition, the only characterization we make for any of the measurements is that the overlap between $G_{B}$ and $F_{B}$ is $\max_{ij}\| F^{i_k}_{B}G^{j_k}_{B}\|^{2}_{\infty}=1/4\;\forall\; k$ (Assumption 4).

If Alice and Bob do one-way classical communication for the post-processing after the protocol from Alice to Bob and Bob's measurement outcomes are used as the raw key (which we call reverse reconciliation), then we can write the Devetak-Winter rate \cite{devetak05} as
\begin{align}
r &\geq H(Z_{B}|E)_{\tau} - H(Z_{B}|Z_A)_{\tau} \label{eq:SDC1} \\
&\geq H(Z_{B}|E)_{\xi} - h_{4}(q_{F}) \label{eq:SDC2}\\
&\geq 2-H(X_{B}|X_{A})_{\sigma} - h_{4}(q_{F}) \label{eq:SDC3}\\
&\geq 2-h_{4}(q_{G}) - h_{4}(q_{F}), \label{eq:SDCkeyrate}
\end{align}
where $h_d$ is the $d$-ary Shannon entropy, $q_F$ is the error rate probability distribution generated from $Z_A$ and $Z_B$, and $q_G$ is the error rate probability distribution generated from $X_A$ and $X_B$. Specifically, these error rate probability distributions consist of the probability that both bits are the same, both bits are different, only the first bit is different, and only the second bit is different.

In going from Eq.~\ref{eq:SDC1} to Eq.~\ref{eq:SDC2} we use the fact that $H(Z_{B}|E)_{\tau}=H(Z_{B}|E)_{\xi}$ and we upper bound the entropy $H(Z_{B}|Z_{A})_{\tau}$ by $h_{4}(q_{F})$ by using the method of types (Lemma II.2 of \cite{csiszar98}). From Eq.~\ref{eq:SDC2} to Eq.~\ref{eq:SDC3} we use the uncertainty relation Eq.~\ref{eq:ur} with the measurements $F_B$ and $G_B$. In Eq.~\ref{eq:SDCkeyrate} we use the method of types to bound $H(X_{B}|X_{A})_{\sigma}$ by $h_{4}(q_{G})$.

Alice and Bob estimate the error rates $q_G$ and $q_F$ by revealing $X_{A}$ and $X_{B}$ as well as a small fraction of their $Z_A$ and $Z_B$ strings in jointly specified positions chosen uniformly at random. Alice and Bob have access to these strings in the prepare and measure SDC protocol because Bob actually performs $F_B$ and $G_B$ (these are the Bell and $Z\otimes X$-measurements in the perfect qubit scenario respectively); Alice uses her encoding bits (which correspond to her string $Z_A$ in the purified protocol via Lemma~1); and her measurement and her resending of the post-measurement state correspond to $X_A$.

We have permutation invariance of the two-bit outcomes and so we can apply the quantum de Finetti theorem of Renner \cite{renner07} to the protocol. Therefore the key rate Eq.~\ref{eq:SDCkeyrate} is applicable for the most general type of attacks by Eve.
Due to the symmetry of the purified protocol we could equivalently do direct reconciliation, where Alice uses her classical string as the key and Bob corrects his raw string. In this case the key rate is the same.

\subsection{LM05 Protocol}

The security proof of this protocol follows the same method as the proof for the SDC protocol, however, there are two differences that need to be taken into account. The first is that Bob chooses a different basis for each of his individual inputs from the channel $Q_2$ according to a classical string, $\Theta$. When a bit of $\Theta$ is $0$, Bob will measure in the $Z$-basis, and when a bit of $\Theta$ is $1$, Bob will measure in the $X$-basis. The other difference is that there are two different measurements that have the desired overlap with Bob's measurement $F_B$ in the uncertainty relation Eq.~1 in the main text. In the perfect purified protocol, Bob's measurement is a $Z\otimes Z$-basis measurement followed by an XOR of the two measurement outcomes. Note that this measurement only has a one bit outcome, and therefore the minimum overlap it can have with another measurement is $1/2$. The $Z\otimes Z$ measurement with an XOR has two measurements with overlap $1/2$, as can be easily verified: measuring the first qubit in the $X$-basis and discarding the second qubit or measuring the second qubit in the $X$-basis and discarding the first qubit.

Now we define three states as we did in the SDC protocol's security proof. We consider the case where reverse reconciliation is performed and we discuss the case of direct reconciliation at the end of this section. The state that Alice and Bob share after they have done their measurements, Bob has publicly announced his basis choices, and Alice has done the sifting of her encoding bits is
\begin{equation}
\tau_{W_{A}W_{B}E} = \sum_{\Theta}F^{\Theta}_{A}\otimes F^{\Theta}_{B} (\rho_{ABE})\otimes\kb{\Theta}{\Theta}.
\end{equation}
The classical outcomes of the measurements for Alice and Bob are written as $W_{A}$ and $W_{B}$ respectively. We assume (Assumption 3) that $F_A^\Theta$ and $F_B^\Theta$ have POVM elements that are independent on each signal so that the uncertainty relation can be applied to each individual measurement. They have the form $\{\bigotimes_{k}F^{\Theta,j_k}_{A}\}_{j}$ and $\{\bigotimes_{k}F^{\Theta,j_k}_{B}\}_{j}$, where $j=j_{1}j_{2}j_{3}\dots$.

For the second state, we change Alice's measurement to be $G^{\Theta,i}_{A}$, which has classical outcome $V^{i}_{A}$, and $i\in\{0,1\}$ is a bit denoting two different measurements Alice could choose. As with the SDC protocol, we do not specify the measurements $G^{\Theta,i}_{A}$. However, we do require that $G^{\Theta,i}_{A}$ has POVM elements of the form $\{\bigotimes_{k}G^{\Theta,i,j_k}_{A}\}_{j}$ (Assumption 3). This gives the state
\begin{equation}
\xi^{i}_{V^{i}_{A}W_{B}E} = \sum_{\Theta}G^{\Theta,i}_{A}\otimes F^{\Theta}_{B} (\rho_{ABE})\otimes\kb{\Theta}{\Theta}.
\end{equation}
Now we also define another state (which we'll use for the uncertainty relation), where we change the measurement on Bob's side to be $G^{\Theta,i}_B$. That is
\begin{equation}
\sigma^{i}_{V^{i}_{A}V^{i}_{B}E} = \sum_{\Theta}G^{\Theta,i}_{A}\otimes G^{\Theta,i}_{B} (\rho_{ABE})\otimes\kb{\Theta}{\Theta}.
\end{equation}
The measurement $G_{B}^{\Theta,i}$ has classical outcome $V^{i}_{B}$. The measurement $G_{B}^{\Theta,i}$ acts independently on each signal, and so its POVM elements have the form $\{\bigotimes_{k}G^{\Theta,i,j_k}_{B}\}_{j}$ (Assumption 3). In addition, $F_B^\Theta$ and $G_{B}^{\Theta,i}$ must satisfy $\max_{jk}\| F^{\Theta,j_l}_{B}G^{\Theta,i,k_l}_{B}\|^{2}_{\infty}=1/2\;\forall\; i,l$ (Assumption 4).

We can now consider the Devetak-Winter rate \cite{devetak05}:
\begin{align}
r &\geq H(W_{B}|E\Theta)_{\tau} - H(W_{B}|W_{A}\Theta)_{\tau} \label{eq:LM051}\\
&\geq H(W_{B}|E\Theta)_{\tau} - H(W_{B}|W_{A})_{\tau} \label{eq:LM052}\\
&\geq H(W_{B}|E\Theta)_{\xi^{i}} - h(q_{F}) \label{eq:LM053}\\
&\geq 1-H(V^{i}_{B}|V^{i}_{A})_{\sigma^{i}} - h(q_{F}) \label{eq:LM054}\\
&\geq 1-h(q_{G^{i}}) - h(q_{F}). \label{eq:LM055}
\end{align}
The error rates $q_{G^i}$ are generated from $V^{i}_{A}$ and $V^{i}_{B}$, and $q_{F}$ is generated from $W_{A}$ and $W_{B}$. Also, the binary entropy is defined as $h(q):=q\log_2 q+(1-q)\log_2 (1-q)$.

From Eq.~\ref{eq:LM051} to Eq.~\ref{eq:LM052} we use the data processing inequality on the second term to trace out $\Theta$. From Eq.~\ref{eq:LM052} to Eq.~\ref{eq:LM053} we use the fact that $H(W_{B}|E\Theta)_{\tau}=H(W_{B}|E\Theta)_{\xi^{i}}$, as well as the method of types to bound the entropy $H(W_{B}|W_{A})$ by $h(q_{F})$ (Lemma II.2 of \cite{csiszar98}). In going from Eq.~\ref{eq:LM053} to Eq.~\ref{eq:LM054} we apply the uncertainty relation Eq.~1 of the main text using the overlap of $1/2$ between the measurements $G_{B}^{\Theta,i}$ and $F_{B}^{\Theta}$. In the last line, Eq.~\ref{eq:LM055}, we use the method of types to bound $H(V^{i}_{B}|V^{i}_{B})_{\sigma^{i}}$ by $h(q_{G^{i}})$. Since Eqs.~\ref{eq:LM051} to \ref{eq:LM055} hold for $i=0$ or $i=1$ we can choose which lower bound on the rate $r$ we would like to use. We would like to have a high lower bound and therefore we pick the minimum of the two binary entropies:
\begin{equation}
r \geq 1-\min_i h(q_{G^{i}}) - h(q_{F}). \label{eq:LM05keyrate}
\end{equation}
To estimate the error rates $q_{G^i}$ and $q_F$ for Version 1 of the LM05 protocol, Alice and Bob reveal $V^i_A$ and $V^i_B$ as well as a small fraction of their $W_A$ and $W_B$ strings in jointly specified positions chosen uniformly at random. In Version 2, Alice and Bob reveal a small fraction of both their $V^i$ strings and $W$ strings in jointly specified uniformly random positions. Alice and Bob have access to these strings in the prepare and measure LM05 protocol because $V^0_A$ is the string of Alice's measurement outcomes and $V^0_B$ is the string of Bob's preparation bits, while $V^1_A$ is the string of Alice's preparation bits (i.e. from her post-measurement state) and $V^1_B$ is the string of Bob's measurement outcomes before doing his XOR. $W_A$ and $W_B$ come from Alice's encoding bit (see Lemma~1), and Bob's XOR of his measurement outcomes and preparation bits. In both versions, the resulting key rate is the same.

It is important to note that since there is a minimization in Eq.~\ref{eq:LM05keyrate} Alice can choose to either not do a measurement or not do a preparation and then the key rate loses the minimization, and instead she just uses the error rate that is estimated ($q_{G^{1}}$ for the former choice and $q_{G^{0}}$ in the latter).

Also, we have permutation invariance of the outcomes (due to the i.i.d. form of the measurements from Assumption 3) and so we can apply the quantum de Finetti theorem of \cite{renner07} to this protocol. Therefore the key rate Eq.~\ref{eq:LM05keyrate} is applicable for the most general type of attacks by Eve. 

In addition, we could have chosen to do direct reconciliation instead of reverse reconciliation. In this case, the string $\Theta$ would represent Alice's choice of encoding from the set $S_0$ or $S_1$. The proof continues in the same manner and the resulting key rate is the same.

\section{Comparison with BB84}

If we set the quantum channels to be fixed resources, then we can use two BB84 protocol implementations to compare with the SDC and LM05 protocols. The first is two one-way BB84 protocols from Alice to Bob (with an asymmetric basis choice so that basis sifting is negligible in the infinite key limit). The second is the Plug \& Play version of BB84 using strong laser pulses (see \cite{gisin02} and references therein). Note that  Plug \& Play BB84 does not have the same level of security as one-way BB84 \cite{zhao08,zhao10}, LM05, or SDC as the measurement devices need to be characterized.
\begin{figure}[h!]
\begin{center}
  \includegraphics[width=8cm]{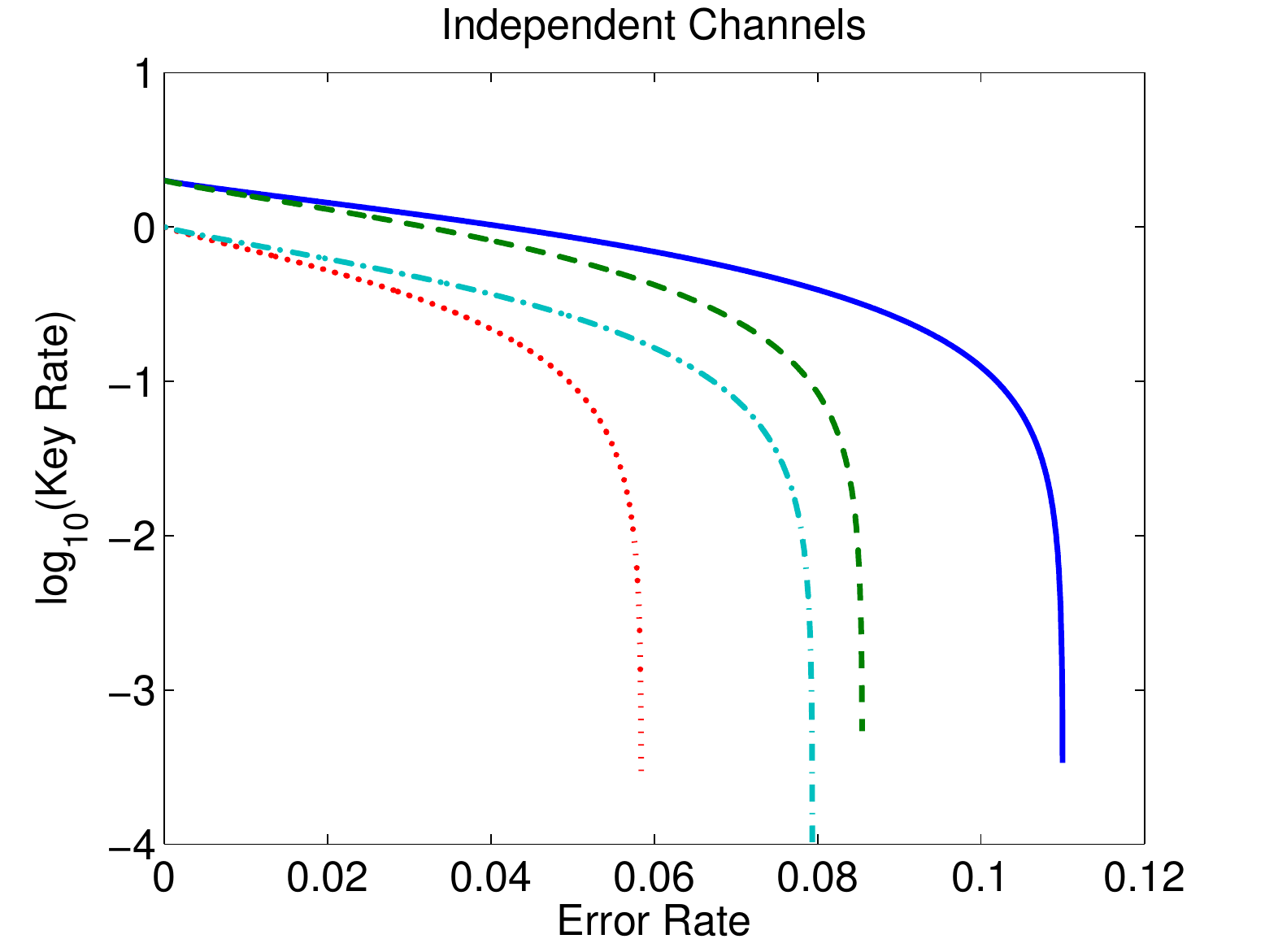}\\
  \vspace{0.2cm}
  \includegraphics[width=8cm]{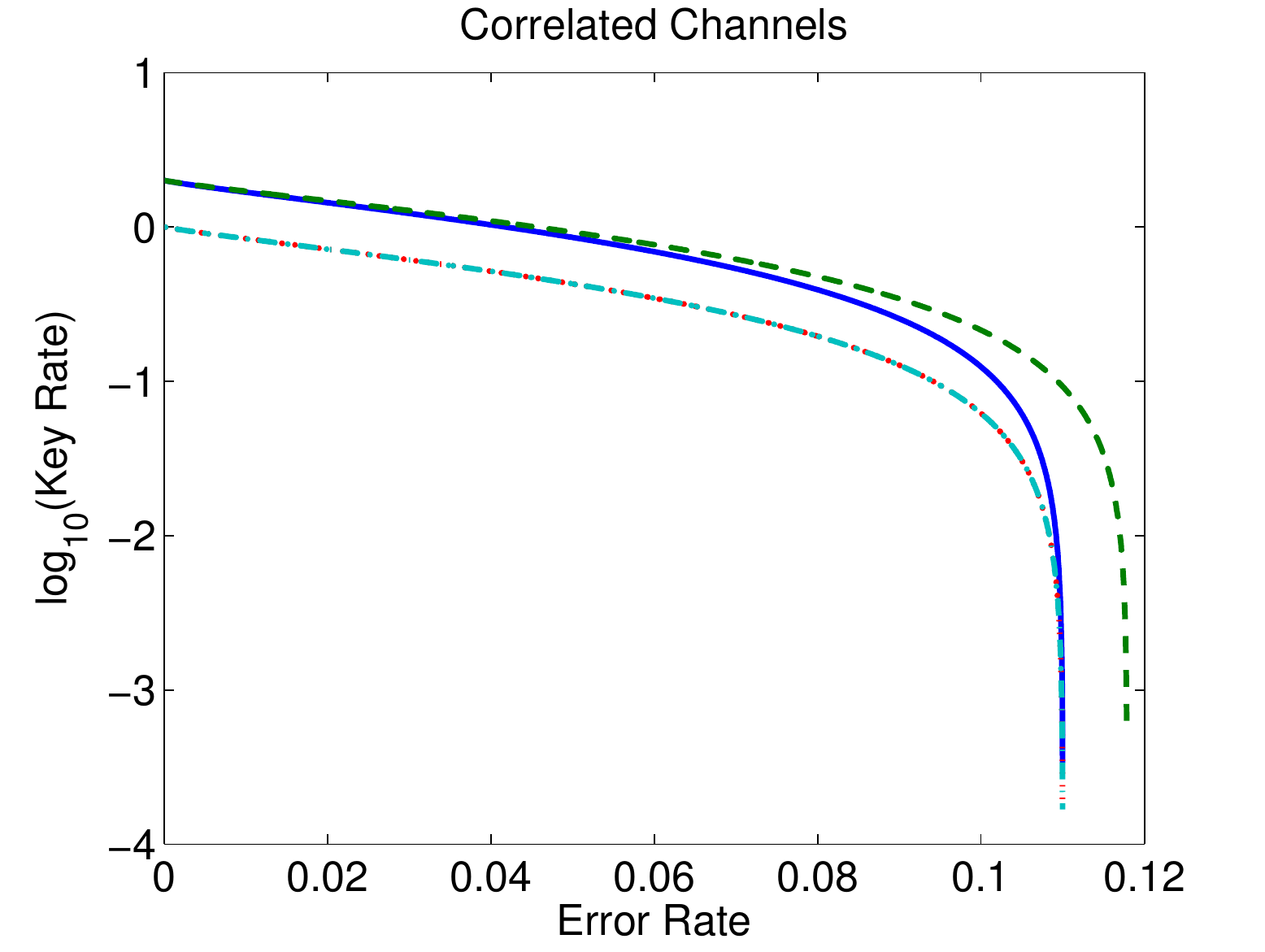}
  \caption{(Color Online) Top: Log base $10$ of the key rates vs.~the error rate (i.e. half the probability of having a state depolarized) for uncorrelated independent identical depolarizing channels.
  Bottom:  Log base $10$ of the key rates vs.~the error rate (i.e. half the probability of having a state depolarized) in one channel, where the channels are correlated such that the probability of becoming depolarized through one channel is the same as the probability of being depolarized when going forwards and backwards through the same channel. The plotted key rates are: two copies of the one-way BB84 protocol performed from Alice to Bob and from Bob to Alice (blue, solid), the SDC protocol (green, dashed), the LM05 protocol (cyan, dot dashed), the Plug \& Play protocol (red, dotted).}\label{fig:graph}
\end{center}
\vspace{-0.8cm}
\end{figure}
If we model the two channels as depolarizing independent identical channels \cite{kumar08,cere06,khir12} 
$\mathcal{E}:\rho\mapsto q\frac{\mathbbm{1}}{d}+(1-q)\rho$,
where $q$ is the probability of depolarizing and
$d$ the dimension of the Hilbert space on which $\rho$ acts, then we see the key rates of Fig.~\ref{fig:graph} (Top). The error rate plotted is $q/2$: the probability of having an error when measuring a signal sent through one of the channels, since with probability $q$ the state is maximally mixed. Since the channels are independent, the probability of being depolarized after passing through both channels in succession is $2q-q^2$.

If instead only one channel is used for communication from Alice to Bob and Bob to Alice, with the polarization drift on the forward channel partially corrected by going back through the channel \cite{lucamarini10}, then the key rates follow Fig.~\ref{fig:graph} (Bottom). That is, the probability of a state being depolarized after passing through the channel is $q$ and the probability of a state being depolarized after passing through the channel one way and then being sent backwards through the same channel is then only $q$ (which is less than $2q-q^2$, which would be the error rate if the channels were independent). In Fig.~\ref{fig:graph} (Bottom) the error rate of the x-axis is also $q/2$ for easy comparison with Fig.~\ref{fig:graph} (Top).

Note that the error rates used to calculate the key rate of the SDC protocol depend upon the probability of getting errors in the first bit only, the second bit only, and both bits of the two-bit measurement outcomes. This means that these error rates for the $G$-measurement basis (the $Z\otimes X$-basis in the perfect implementation) are $q/2(1-q/2), (1-q/2)q/2$, and $q^2/4$ respectively for the situations in Fig.~\ref{fig:graph}. For the $F$-measurement basis (the Bell-basis measurement in the perfect implementation) the error rates are all $(2q-q^2)/4$ for the situation in Fig.~\ref{fig:graph} (Top) and $q/4$ for the situation in Fig.~\ref{fig:graph} (Bottom).

Importantly, the SDC protocol key rate exceeds both BB84 key rates in the scenario of Fig.~\ref{fig:graph} (Bottom), and it can also tolerate a higher error rate of $11.8\%$. 
This is because the correlation between the forward and backward channel makes the error rate in the $F$-measurement basis lower. This advantage increases if the error rate of passing forwards and then backwards through the channel is smaller.

\section{Conclusion}

We have shown a general method to prove security of two-way QKD protocols. We have applied this proof method to two such protocols, namely one based on super dense coding (SDC), and another based on a previously proposed two-way protocol (LM05) \cite{lucamarini05}. These two protocols are secure against the most general types of attacks by an eavesdropper and provide the following key rates:
\begin{equation}
\begin{array}{clcc}
\textrm{SDC:} & r_{SDC} \geq 2-h_{4}(q_{G}) - h_{4}(q_{F}) \\
\textrm{LM05:} & r_{LM05} \geq 1-\min_{i}h(q_{G^{i}}) - h(q_{F}),
\end{array}
\end{equation}
where in the later $i=0,1$ denotes two possible measurements Alice could choose.
Importantly, few assumptions are needed about the devices used. This is a step towards device independence for two-way QKD protocols. We make the following assumptions to apply our security proof: preparations are done in a purified way (i.e. an arbitrary bipartite state is prepared and half of it is measured, while the other half is used as the preparation), Alice's encoding output is a fixed state, measurements are done independently on each signal, and a fixed overlap constant characterizes either Bob or Alice's devices (depending on whether reverse or direct reconciliation is performed). The first assumption can instead be the assumption that qubits are prepared. Interesting future work could be to remove some of these assumptions while still providing the same rates of security.

We have shown that these protocols have comparable performance to different implementations of the BB84 protocol, and can even exceed the BB84 rate in certain relevant parameter regimes. In addition, the key rate we obtain for the LM05 protocol is higher than that of \cite{lu11}.

The determinism of two-way protocols in the infinite-key case is not an advantage since an asymmetrical basis choice in the BB84 protocol makes it deterministic as well. However, in the finite-key regime, the BB84 protocol is not deterministic \cite{tomamichel12a}. Therefore both the SDC and LM05 protocols will have an advantage over BB84 implementations when finite keys are used.

In addition, an advantage that the LM05 protocol has, which is not apparent in the infinite key limit, is that there is a higher fraction of key bits per signal sent compared to the BB84 and SDC protocols. If the basis bias for BB84 and the SDC protocol used for parameter estimation is $p$, then \mbox{$2p(1-p)$} fraction of the signals are lost due to basis sifting. However, in the LM05 protocol, if $c$ is the probability that Alice does her measurement, and $p$ is the probability that Alice and Bob use the $Z$-basis, then only $2p(1-p)c$ fraction of the signals are lost. This advantage would have a positive effect on the finite-key rate.

Our work paves the way for fully exploiting the potential of entropic uncertainty relations in two-way QKD with finite-key sizes for any possible implementation. We did not evaluate the finite-key regime here, but the techniques of \cite{tomamichel11a,tomamichel12a} could be used to show security for two-way protocols. We leave this as future work.

\acknowledgments
\emph{Acknowledgments:}
The authors thank J. {\AA}berg, F. Dupuis, B. Fortescue, F. Fung, H.-K. Lo, N. L{\"u}tkenhaus, X. Ma, B. Qi,  and J. Renes for helpful discussions and insight. M.L. and S.M. are also grateful to ETH for kind hospitality during the early stages of this work. N.J.B. is grateful to the Universit{\` a} di Camerino for hospitality where part of this work was completed. Part of M.L.'s work has been done under the 5{\textperthousand} grant C.F. 81001910439. This work was supported by SNSF through the National Centre of Competence in Research ÒQuantum Science and TechnologyÓ and through grant No. 200020-135048, and by the European Research Council through grant No. 258932.

\bibliographystyle{ieeetr}

\begin{thebibliography}{90}

\bibitem{scarani09}
V.~Scarani, H.~Bechmann-Pasquinucci, N.~Cerf, M.~Du\v{s}ek, N.~L\"{u}tkenhaus,
  and M.~Peev, Rev. Mod. Phys. {\bf 81}, 1301 (2009).

\bibitem{gisin02}
N.~Gisin, G.~Ribordy, W.~Tittel, and H.~Zbinden, Rev. Mod. Phys. {\bf 74}, 145 (2002).

\bibitem{cai04a}
Q.-y.~Cai and B.-w.~Li, Phys. Rev. A {\bf 69}, 054301 (2004).

\bibitem{cai04b}
Q.-y.~Cai and B.-w.~Li, Chin. Phys. Lett. {\bf 21}, 601 (2004).

\bibitem{deng04}
F.-G.~Deng and G.L.~Long, Phys. Rev. A {\bf 69}, 052319 (2004).

\bibitem{bostrom02}
K.~Bostr\"{o}m and T.~Felbinger, Phys. Rev. Lett. {\bf 89}, 187902 (2002).

\bibitem{lucamarini05}
M.~Lucamarini and S.~Mancini, Phys. Rev. Lett. {\bf 94}, 140501 (2005).

\bibitem{cere06}
A.~Cer\`{e}, M.~Lucamarini, G.~{Di Giuseppe}, and P.~Tombesi, Phys. Rev. Lett. {\bf 96}, 200501 (2006).

\bibitem{kumar08}
R.~Kumar, M.~Lucamarini, G.~{Di Giuseppe}, R.~Natali, G.~Mancini, and
  P.~Tombesi, Phys. Rev. A {\bf 77}, 022304 (2008).

\bibitem{ostermeyer08}
M.~Ostermeyer and N.~Walenta, Opt. Commun. {\bf 281}, 4540 (2008).

\bibitem{khir12}
M.~{Abdul Khir}, M.~{Mohd Zain}, I.~Bahari, and S.~Shaari, Opt. Commun. {\bf 285}, 842 (2012).

\bibitem{bennett92b}
C.H.~Bennett and S.J.~Wiesner, Phys. Rev. Lett. {\bf 69}, 2881 (1992).

\bibitem{yuan05}
Z.~L. Yuan and A.~J. Shields, Opt. Exp. {\bf 13}, 660 (2005).

\bibitem{dixon10}
A.~R. Dixon, Z.~L. Yuan, J.~F. Dynes, A.~W. Sharpe, and A.~J. Shields, Appl. Phys. Lett. {\bf 96}, 161102 (2010).

\bibitem{marand95}
C.~Marand and P.~D. Townsend, Opt. Lett. {\bf 20}, 1695 (1995).

\bibitem{bb84}
C.~H. Bennett and G.~Brassard, in {\em Proceedings of IEEE International Conference on
  Computers, Systems and Signal Processing, Bangalore, India, 1984, (IEEE, New York, 1984)} p.~175.

\bibitem{ribordy98}
G.~Ribordy, J.-D. Gautier, N.~Gisin, O.~Guinnard, and H.~Zbinden, Electr. Lett. {\bf 34}, 2116 (1998).

\bibitem{wojcik03}
A.~W\'{o}jcik, Phys. Rev. Lett. {\bf 90}, 157901 (2003).

\bibitem{cai03}
Q.-y. Cai, Phys. Rev. Lett. {\bf 91}, 109801 (2003).

\bibitem{lu11}
H.~Lu, C.-H.~F.~Fung, X.~Ma, and Q.-y. Cai, Phys. Rev. A {\bf 84}, 042344 (2011).

\bibitem{fung12a}
C.-H.~F.~Fung, X.~Ma, H.~F. Chau, and Q.-y. Cai, Phys. Rev. A {\bf 85}, 032308 (2012).

\bibitem{zhao08}
Y.~Zhao, B.~Qi, and H.-K. Lo, Phys. Rev. A {\bf 77}, 052327 (2008).

\bibitem{zhao10}
Y.~Zhao, B.~Qi, H.-K. Lo, and L.~Qian, New J. Phys. {\bf 12}, 023024 (2010).

\bibitem{mayers96}
D.~Mayers, {\em Proceedings of Advances in Cryptology -- CRYPTO '96, Santa Barbara, California, USA, 1996, Lecture Notes in Computer Science (Springer, New York, 1996),} p.~343.

\bibitem{koashi09}
M.~Koashi, New J. Phys. {\bf 11}, 045018 (2009).

\bibitem{barrett05}
J.~Barrett, L.~Hardy, and A.~Kent, Phys. Rev. Lett. {\bf 95}, 010503 (2005).

\bibitem{barrett12}
J.~Barrett, R.~Colbeck, and A.~Kent, Phys. Rev. A {\bf 86}, 062326 (2012).

\bibitem{vazirani12}
U.~Vazirani and T.~Vidick, arXiv:1210.1810.

\bibitem{berta10}
M.~Berta, M.~Christandl, R.~Colbeck, J.~M. Renes, and R.~Renner, Nat. Phys. {\bf 6}, 659 (2010).

\bibitem{devetak05}
I.~Devetak and A.~Winter, Proceedings of the Royal Society A {\bf 461}, 207 (2005).

\bibitem{tomamichel12a}
M.~Tomamichel, C.~C.~W. Lim, N.~Gisin, and R.~Renner, Nat. Comm. {\bf 3}, 634 (2012).

\bibitem{tomamichelthesis}
M.~Tomamichel, Ph.D. thesis, ETH Zurich, arXiv:1203.2142, 2012.

\bibitem{CHSH}
J.~Clauser, M.~Horne, A.~Shimony, and R.~Holt, Phys. Rev. Lett. {\bf 23}, 880 (1969).

\bibitem{tomamichel13}
M.~Tomamichel and E.~H\"{a}nggi, J. of Phys. A {\bf 46}, 055301 (2013).

\bibitem{csiszar98}
I.~Csiszar, IEEE Trans. on Inf. Theory {\bf 44}, 2505 (1998).

\bibitem{renner07}
R.~Renner, Nat. Phys. {\bf 3}, 645 (2007).

\bibitem{lucamarini10}
M.~Lucamarini and S.~Mancini, arXiv:1004.0157.

\bibitem{tomamichel11a}
M.~Tomamichel and R.~Renner, Phys. Rev. Lett. {\bf 106}, 110506 (2011).

\bibitem{choi75}
M.-D. Choi, Lin. Alg. and its Appl. {\bf 10}, 285 (1975).

\bibitem{jamiolkowski72}
A.~Jamiolkowski, Rep. on Math. Phys. {\bf 3}, 275 (1972).

\end{thebibliography}

\section{Appendix}

Here we provide the proof of Lemma~1 that purifies Alice's encoding operation. It establishes an equivalence between a POVM acting on half of a pure state and a CPTP map of a particular form.

\begin{lemma1}[POVM equivalent to a CPTP map]\label{lemma:POVMCPTP} Let $\{\mathcal{E}_{i}\}_{i=1..n}$ be a set of $n$ completely positive trace-preserving maps from Hilbert space $\mathcal{H}_A$ to Hilbert space $\mathcal{H}_D$ and $\sigma_{D}$ be a fixed density operator on $\mathcal{H}_{D}$ such that $1/n\sum_{i=1}^{n}\mathcal{E}_{i}(\rho_{A})=\sigma_{D}\;\forall \rho_{A}\in S(\mathcal{H}_{A})$.

Then there exists a fixed pure state $\ket{\phi}_{CD}$  in $\mathcal{H}_{CD}:=\mathcal{H}_{C}\otimes\mathcal{H}_{D}$, where $\dim \mathcal{H}_{C} = \dim\mathcal{H}_{D}$, and a complete set of POVM elements $\{F^{i}_{AC}\}_{i=1..n}$ on $\mathcal{H}_{AC}$ $($so $\sum_{i}F^{i}_{AC}=\mathbbm{1}_{AC})$, such that $\forall i,\forall\rho_{A}\in S(\mathcal{H}_{A})$ we have
\begin{equation}\label{eq:equiv1}
n\Tr_{AC}\left(F_{AC}^{i}\rho_{A}\otimes\ket{\phi}_{CD}\bra{\phi}\right)=\mathcal{E}_{i}(\rho_{A}).
\end{equation}
\end{lemma1}
\begin{proof}
Summing over $i$ in Eq.~\ref{eq:equiv1} implies that we require $\Tr_{C}(\ket{\phi}_{CD}\bra{\phi})$=$\sigma_{D}$, and therefore we fix $\ket{\phi}_{CD}$ to be a purification of $\sigma_{D}$. Now we can constructively determine what the POVM elements $F^{i}_{AC}$ are in terms of $\sigma_{D}$ and the maps $\mathcal{E}_{i}$. Then we will show that this construction of the POVM satisfies all necessary requirements above.

Let $\sigma_{D}=\sum_{j}\lambda_{j}\ket{j}_{D}\bra{j}$, so then $\ket{\psi}_{CD}=\sum_{j}\sqrt{\lambda_{j}}\ket{jj}_{CD}$. Expanding $\rho_{A}$ in an orthonormal basis $\{\ket{\psi_{m}}\}_{m}$ gives $\rho_{A}=\sum_{ml}r_{ml}\ket{\psi_m}_{A}\bra{\psi_l}$, which allows us to write Eq.~\ref{eq:equiv1} as
\begin{align}
&\sum_{jkml}n\Tr_{AC}\left(F_{AC}^{i}r_{ml}\ket{\psi_m}_{A}\bra{\psi_l}\otimes\sqrt{\lambda_{j}\lambda_{k}}\ket{jj}_{CD}\bra{kk}\right) \nonumber\\
&=\sum_{jkml}nr_{ml}\sqrt{\lambda_{j}\lambda_{k}} \left(_{AC}\bra{\psi_l k}F_{AC}^{i}\ket{\psi_m j}_{AC}\ket{j}_{D}\bra{k}\right) \nonumber\\
&=\sum_{ml}r_{ml}\mathcal{E}_{i}(\ket{\psi_m}_{A}\bra{\psi_l}).
\end{align}
This must be true for all $\rho_{A}$ and therefore we have $\forall\; m,l$
\begin{align}
&\sum_{jk}n\sqrt{\lambda_{j}\lambda_{k}} \bra{\psi_l k}F_{AC}^{i}\ket{\psi_m j}\ket{j}_{D}\bra{k} =\mathcal{E}_{i}(\ket{\psi_m}\bra{\psi_l}),\nonumber\\
&n \sqrt{\lambda_{j}\lambda_{k}}\bra{\psi_l k}F_{AC}^{i}\ket{\psi_m j} = \bra{j}\mathcal{E}_{i}(\ket{\psi_m}\bra{\psi_l})\ket{k} \;\forall m,l,j,k. \label{eq:coeffs}
\end{align}
Eq.~\ref{eq:coeffs} gives a constructive way of finding the POVM elements $F_{AC}^{i}$. If $\sigma_{D}$ has full rank then $F_{AC}^{i}$ is completely determined by this equation. If $\sigma_{D}$ is not of full rank then $F_{AC}^{i}$ can be decomposed into a part on the support of $\sigma_{C}:=\Tr_{D}(\ket{\phi}_{CD}\bra{\phi})$ and its kernel: $F_{AC}^{i}=F_{AC_{\supp \sigma_{C}}}^{i}\oplus F_{AC_{\text{kern} \sigma_{C}}}^{i}$. The block on the $\supp\sigma_{C}$ is completely specified by Eq.~\ref{eq:coeffs}, and the block on $\text{kern}\sigma_{C}$ can be chosen arbitrarily as long as $F_{AC_{\text{kern} \sigma_{C}}}^{i}\geq 0,$ for all $i$ and satisfy $\sum_{i}F_{AC_{\text{kern} \sigma_{C}}}^{i}=\mathbbm{1}_{AC_{\text{kern} \sigma_{C}}}$. It is clear from Eq.~\ref{eq:coeffs} that $\sum_{i}F_{AC_{\supp\sigma_{C}}}^{i}=\mathbbm{1}_{AC_{\supp\sigma_{C}}}$.

Now we need to verify that the POVM elements satisfy $F_{AC}^{i}\geq 0$ for all $i$. We write the maps in their Choi-Jamio\l kowski representation \cite{choi75,jamiolkowski72}:
\begin{align}
\mathcal{E}_{i}(\ket{\psi_m}_{A}\bra{\psi_l})&= \Tr_{A}(J_{AD}^{i}(\ket{\psi_m}_{A}\bra{\psi_l})^{T}\otimes\mathbbm{1}_{D}) \\
&=\bra{\psi_m}J^{i}_{AD}\ket{\psi_l},
\end{align}
where $J^{i}_{AD}$ are the Choi-Jamio\l kowski matrices for the maps $\mathcal{E}_{i}$.
Now we can write $F_{AC_{\supp \sigma_{C}}}^{i}$ from Eq.~\ref{eq:coeffs} as
\begin{equation}
F_{AC_{\supp \sigma_{C}}}^{i} = \frac{1}{n}\frac{1}{\sqrt{\sigma_{C}}}(J^{i}_{AC})^{T}\frac{1}{\sqrt{\sigma_{C}}},
\end{equation}
where $J^{i}_{AC}:=\sum_{jk}\ket{j}_{CD}\bra{j}J^{i}_{AD}\ket{k}_{DC}\bra{k}$. From this form it is clear that this block is positive, and so $F_{AC}^{i}\geq 0$ for all $i$.
\end{proof}

\end{document}